\documentclass{article}

\usepackage{amsmath}

\newcommand{\be}{\begin{equation}}
\newcommand{\ee}{\end{equation}}
\newcommand{\ben}{\begin{eqnarray}}
\newcommand{\een}{\end{eqnarray}}

\newcommand{\nd}{\noindent}

\title{{\bf Tsallis' Statistics without probability distributions}\\  REPLY to Comment on \\ {\bf Possible Divergences in Tsallis' Thermostatistics}}
\author{A. Plastino, M. C. Rocca\\
La Plata National University\\
and Argentina's National Research Council,\\
(IFLP-CCT-CONICET)-C. C. 727, 1900\\
La Plata - Argentina}

\date{\today}

\begin{document}

\maketitle

\begin{abstract}

\nd In a recent letter (EPL, 104 (2013) 60003) we suggested a way
to avoid divergences  inherent to the formulation of nonextensive
statistical mechanics. They can be eliminated via the use of a
q-Laplace transformation, which was illustrated for the case of
the harmonic oscillator. Lutsko and Boon now contend, in an
interesting comment, that our  new formulation raises questions
that they carefully discuss. Here we elaborate an explanation of
the contents of (EPL, 104 (2013) 60003)  that permits to
appreciate our original Letter in a more positive fashion.

\vskip 3mm  \nd PACS {05.20.-y, 05.70.Ce, 05.90.+m}

\end{abstract}

\nd It is well known that, for obtaining the partition function
$\cal{Z}$, two alternative routes can be followed:

\begin{itemize}

\item the ``natural" one, given by $\cal{Z}$'s definition in terms of a sum over ``un-normalized" probabilities, and

\item $\cal{Z}$   as the Laplace Transform of the energy density.

\end{itemize}

\nd In the orthodox Boltzmann-Gibbs Statistical Mechanics, that
uses the ordinary exponential function, the two routes yield the
same result.\vskip 3mm

\nd We proved in \cite{uno}  that such is NOT the case for
Tsallis' thermostatistics, for which the first alternative (use of
a probability distribution) {\sf diverges} in one or more
dimensions, due to the long tail of the q-exponential function.
One must necessarily follow the second path (without employing
probability distributions), that yields finite results. Thus, the
q-Laplace Transform is seen to become an indispensable tool for
nonextensive statistics.

\nd In their Comment, Lutsko and Boon (LB) raise four interesting
points (their essence given below in italics)) that certainly
deserve detailed debate.

\begin{enumerate}

\item LB assert that, {\it while our procedure yields a finite value for the partition function), the original q-exponential distribution  remains
un-normalizable}. True, but our whole point is that we do NOT want
to employ probability distributions (PD) in out treatment. Our
only microscopic input is the energy-density, as discussed, for
example, in Reif's text-book \cite{reif}.

\item {\it Our formalism would not satisfy the relationship $U=-(\partial \beta F/\partial
\beta)_V$.} The $f-$definition above is wrong. We know that
\[S_q=k Z_q^{q-1}\left(\ln_q Z_q+\beta <U_q>\right),\]
where $l$ stands for Boltzmann's constant. Using now the
presription:
\[F_q=<U_q>-TS_q,\]
we find
\[F_q=(1-Z_q^{q-1})<U_q>-\frac {Z_q^{q-1}} {\beta}\ln_q Z_q,\]
an expression that does not coincide with that of LB for $F_q$.
From our last relation one finds
\[\frac {\partial\beta F_q} {\partial \beta}=
(1-Z_q^{q-1})<U_q>+\beta (1-Z_q^{q-1})
\frac {\partial<U_q>} {\partial\beta}-
\beta<U_q>\frac {\partial Z_q^{q-1}} {\partial\beta}-\]
\[\frac {\partial Z_q^{q-1}} {\partial\beta} \ln_q Z_q-
Z_q^{q-1}\frac {\partial\ln_q Z_q} {\partial\beta},\] that, for
$q=1$, reduces to
\[\frac {\partial\beta F} {\partial\beta}=
-\frac {1} {Z} \frac {\partial Z} {\partial\beta}=<U>,\] in full
agreement with Tsallis' prescription. The evaluation of  $\frac
{\partial\beta F_q} {\partial \beta}$ is a function of $q$ that
turns out to coincide with  $<U>$ for  $q=1$, which, in turn,
contradicts LB`s assertions.

\item  {\it Now the resulting entropy is not equivalent to
the original Tsallis entropy evaluated with the q-exponential
distribution} (as probability distributions). Of course it is not!
We do away with probability distributions in order to avoid the
Tsallis divergences.

\item  {\it The expansion used in Eq.(4) of \cite{uno}  seems quite arbitrary. One could, for example,
replace $a_n x^n by$ $2^n a_n) (x/2)^n$ and thereby obtain an
inequivalent form}. Our answer is that the Mc Laurin expansion is
UNIQUE and can not be arbitrarily modified as LB want.

\end{enumerate}

\nd Summing up, the LB Comments are interesting and stimulate
fertile debate around mathematical niceties the enter Tsallis'
theory. These certainly need to be fully explored. Their questions
and our answers will hopefully serve such purpose.

\section*{Acknowledgements}

\nd The authors acknowledge support from CONICET (Argentine
Agency).

\end{document}